# Generic Optimization of Linear Precoding in Multibeam Satellite Systems

Gan Zheng, Symeon Chatzinotas, and Björn Ottersten

**Abstract**

Multibeam satellite systems have been employed to provide interactive broadband services to geographical areas under-served by terrestrial infrastructure. In this context, this paper studies joint multiuser linear precoding design in the forward link of fixed multibeam satellite systems. We provide a generic optimization framework for linear precoding design to handle any objective functions of data rate with general linear and nonlinear power constraints. To achieve this, an iterative algorithm which optimizes the precoding vectors and power allocation alternatingly is proposed and most importantly, the proposed algorithm is proved to always converge. The proposed optimization algorithm is also applicable to nonlinear dirty paper coding. As a special case, a more efficient algorithm is devised to find the optimal solution to the problem of maximizing the proportional fairness among served users. In addition, the aforementioned problems and algorithms are extended to the case that each terminal has multiple co-polarization or dual-polarization antennas. Simulation results demonstrate substantial performance improvement of the proposed schemes over conventional multibeam satellite systems, zero-forcing and regularized zero-forcing precoding schemes in terms of meeting the traffic demand. The performance of the proposed linear precoding scheme is also shown to be very close to the dirty paper coding.

**Keywords**

(Multibeam Satellite, precoding, optimization, dual-polarization)

## I. Introduction

Multibeam satellite systems have been inspired by the success of the cellular paradigm, which allows carefully planned frequency reuse while keeping intercell interference within acceptable limits to achieve high spectral efficiency. In addition, the demand for interactive data services on top of broadcasting has supported the implementation of multibeam systems, which allow for finer partitioning of the coverage area and independent stream transmission within each beam.

A large number of spotbeams can be employed to cover the same coverage area contrary to recent satellite technology where a single (global) beam is employed. Currently, tens or hundreds of beams are possible with a typical reuse factor of four. However, due to the antenna design, the beam patterns partially

The authors are with the Interdisciplinary Centre for Security, Reliability and Trust (SnT), The University of Luxembourg, 6 rue Richard Coudenhove-Kalergi, L-1359 Luxembourg-Kirchberg, Luxembourg. Email: {gan.zheng, symeon.chatzinotas, bjorn.ottersten}@uni.lu.

Björn Ottersten is also with the Signal Processing Laboratory, ACCESS Linnaeus Center, KTH Royal Institute of Technology, SE-100 44 Stockholm, Sweden. Email: bjorn.ottersten@ee.kth.se.

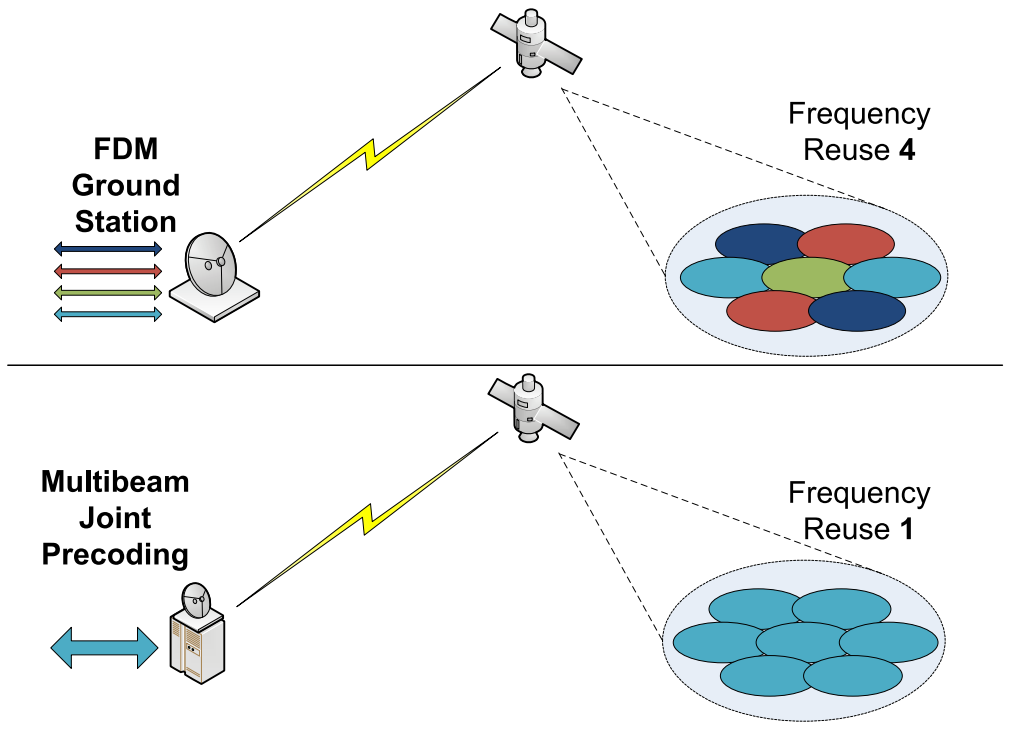

Fig. 1. A multibeam satellite system with a Ground Station serving a 7-beam cluster. Top: Conventional 4-color frequency reuse scheme. Bottom: Multibeam joint precoding paradigm using full frequency reuse.

overlap on the ground creating interbeam interference. The beam patterns and the corresponding allocated power have to be carefully designed to ensure that interbeam interference stays within acceptable limits, which are determined by the carrier to interference ratio of the beam-edge users. A similar effect has been limiting the performance of terrestrial cellular networks for decades, but has been alleviated based on multicell joint processing, where user signals in the downlink channel are jointly precoded before being transmitted by neighboring BS antennas in order to mitigate inter-cell interference. However, one of the practical obstacles in terrestrial implementation is the requirement of a backhaul network which enables the cooperation amongst neighboring BSs.

The principle of multibeam joint processing can be applied to multibeam satellite systems. As illustrated in Fig. 1, instead of being served by only one beam, each user's signal is precoded at the gateway (GW) and sent by all beams. The main implementation advantage over terrestrial wireless systems is that usually the signals for adjacent beams are transmitted from the same GW through the satellite to the users in the forward link (FL), as a result, joint precoding can take place at that GW and there is no need for expensive backhauling. When multiple GWs serve clusters of beams, distributed joint precoding techniques can be employed, but in this paper we focus on one cluster of beams served by a single GW. To mitigate interference among multibeams, spatial processing and specifically effective precoding techniques can be exploited, which jointly pre-processes data to all beams at the GW. It was reported in [1] that with simple linear precoding, an improvement of the achievable spectral efficiency of about $25 \sim 50\%$ can be achieved.

In the following, we provide a review on the available multiuser joint precoding techniques in terrestrial wireless communications, as well as an overview of related work in the satellite literature.

### A. *Precoding Techniques and Satellite Literature*

In the multiple-input multiple-output (MIMO) and multiple-input single-output (MISO) FL literature, a number of linear and non-linear techniques have been proposed to effectively manage interference among users. Dirty paper coding (DPC) is an optimal non-linear technique based on known interference pre-cancellation which has been shown to achieve the MIMO downlink sum capacity [2] [3]. Although DPC is the optimal capacity-achieving transmission strategy, it is not suitable for practical implementation and mostly serves as a performance upper bound for the evaluation of less complex techniques. Tomlinson-Harasima precoding (THP) is another nonlinear precoding technique which is a more practical implementation of DPC based on modulo operations over the constellation symbols [4]. To further reduce the implementation complexity, linear precoding schemes have received much attention. Among them, zero-forcing (ZF) is one of the simplest techniques that prefilters the transmit signal vector with the channel pseudo inverse [5] so that there exists no inter-user interference. Regularized ZF (R-ZF) [6] extends ZF by taking into account the noise variance in order to improve performance in the low SNR regime. Opportunistic beamforming (OB) is another linear technique, where each user selects amongst predefined random beamformers from a codebook based on both instantaneous channel state information (CSI) and long-term channel statistics [7].

In the literature of satellite communications, the majority of early work focuses on the conventional scenario where polarization or partial frequency reuse is employed to mitigate interbeam interference. For example, authors in [8] optimize the power and beam allocation in order to adapt to channel conditions and meet traffic demands. Recently, multibeam joint processing scenarios have been studied in various settings. To be specific, FL cases have been investigated in [1], [10]–[13], while reverse link cases in [14]–[16]. In all FL studies, fixed satellite services were considered since reliable CSI feedback can only be acquired for slow-fading channels due to the long propagation delays. Various characteristics of the multibeam satellite channel were taken into account such as beam gain [9], [12], [13], rain fading [1], interference matrix [12] and correlated attenuation areas [13]. In terms of precoding techniques, THP was studied in [9] and [12], while linear precoding such as ZF and R-ZF were evaluated in [1], [9]. Finally, authors in [13] have considered an OB technique based on a codebook of orthonormal precoders and low-rate feedback.

### *B. Contributions*

In previous work, optimized linear precoding design with concrete objectives and constraints has not been treated therefore it is hard to evaluate the potential of each precoding scheme. This paper aims to fill this gap and studies the linear precoding design for the FL of a fixed multibeam satellite system with perfect CSI using optimization techniques. We take into account the power flexibility, which is essential for optimum resource allocation in multibeam satellite communications. It can be implemented by using traveling wave tube amplifiers (TWTAs) [22] based on the interaction between an electron beam and the radio wave within a tube and the multi-port amplifiers (MPAs) [23], where the total available power of a set of amplifiers can be flexibly distributed amid different beams. We first assume the user terminals employ single polarization then extend it to the case of dual polarization, which correspond to multiuser MISO and multiuser MIMO systems, respectively.

The design objective is to optimize any given function of individual users' rates subject to general linear and nonlinear power constraints which can model the power flexibility. Specifically, this paper makes the following contributions:

1) We formulate a general precoding design problem by incorporating any functions of individual users'

rate and general linear and nonlinear power constraints [1].

2) We propose a generic iterative algorithm based on alternating optimization to tackle the above problem and most importantly prove its convergence. We also devise an efficient algorithm for a sub-problem of power minimization. The proposed algorithm is applicable to nonlinear DPC precoding.
3) We extend the proposed algorithm to cope with the case where satellite terminals have multiple receive dimensions, and one particular important scenario is when two co-polarization or cross-polarization antennas are used at receivers. Polarization correlation and discrimination are considered in the channel modeling.

The main findings via simulations are summarized below

i) The performance of the optimized linear precoding is very close to that of the optimized nonlinear DPC precoding, which provides a performance upper bound;
ii) To improve the performance, the flexible transmit power enabled by TWTAs and MPAs is even more important than the nonlinear precoding.

The optimization indeed adds to the complexity of signal processing, but this is not a major issue since the precoding design is optimized at the GW and thus no additional computation is needed on the satellite.

The remainder of this paper is structured as follows. In section II, we introduce the fixed multibeam satellite system model and the problem of linear precoding design with general objectives and general linear and nonlinear power constraints. In Section III, we devise a generic iterative algorithm by optimizing power allocation and precoding vectors alternatingly and prove its convergence. Section IV modifies the proposed algorithm to optimize DPC precoding with fixed encoding order and also as a special case, we propose an algorithm to find the optimal solution for the fairness maximization problem with convex power constraints. In Section V, the proposed algorithm is extended to deal with the case when satellite terminals employ co-polarization or cross-polarization. In Section VI, the proposed multibeam precoding is compared to conventional multibeam systems and existing precoding techniques through numerical simulations and Section VII concludes the paper.

### C. Notations

Throughout this paper, the following notations will be adopted. $|\cdot|$ denotes the modulus of a complex scalar. Vectors and matrices are represented by bold lowercase and uppercase letters, respectively, and $\|\cdot\|$ is the Frobenius norm. The superscript $^\dagger$ is used to denote the Hermitian transpose of a vector or matrix. $\mathbf{A} \succeq \mathbf{0}$ means that matrix $\mathbf{A}$ is positive semi-definite. $\mathbf{A} \odot \mathbf{B}$ and $\mathbf{A} \otimes \mathbf{B}$ denote Hadamard product and Kronecker product of two matrices, respectively. $\mathbf{I}$ denotes an identity matrix. $\mathbb{I}_n$ and $\mathbf{1}_n$ denote an $n \times n$ all-one matrix and an $n \times 1$ vector, respectively. $\mathbb{E}[\cdot]$ denotes the expected value of a random variable. Finally, $\mathbf{x} \sim \mathcal{CN}(\mathbf{m}, \mathbf{\Theta})$ denotes a vector $\mathbf{x}$ of complex Gaussian entries with a mean vector of $\mathbf{m}$ and a covariance matrix of $\mathbf{\Theta}$.

[1]There are mild conditions about general cost functions and nonlinear power constraints which are satisfied by all reasonable performance metrics and will be discussed later.

## II. System Model

Consider a FL bent-pipe satellite system for transmitting independent streams to multiple fixed terminals through multiple beams. A single GW who has perfect CSI manages a cluster of $K$ adjacent beams on ground formed by $K$ antenna feeds (single-feed per beam) on board the satellite and full frequency reuse among beams is assumed. This system resembles a multiuser MISO downlink in terrestrial communications. By employing a time division multiplexed (TDM) scheme, a single user per beam is served for each time slot. The uplink of the feeder link is ideal and no inter-cluster interference is considered. We also assume the channel experiences slow fading and perfect CSI is available at the GW. This can be realized by feedback/training sent from the terminals via return a return channel, which already exists in DVB-S2 [24].

### A. Satellite Channel Model

One of the main reasons why satellite communications are challenging and different from terrestrial communications is due to the satellite channel characteristics, which need to be properly modeled. The satellite channel above 10 GHz operating under line-of-sight (LOS) is subjected to various atmospheric fading effects originating in the troposphere, which severely degrade system performance and availability [17]. Among them, rain attenuation is the dominant factor and will be taken into account in our modeling. In the following we will describe in detail the satellite channel effects including free space loss, rain fading and the beam gain pattern.

*1) Free Space Loss (FSL):* Due to the earth curvature and the wide satellite coverage, the free space loss in each multibeam will not be identical. In order to model this effect, the FSL coefficient of the $k$th multibeam can be written as [18]:

$$b_{\max}(k) = \left(\frac{\lambda}{4\pi}\right)^2 \frac{1}{{d_0}^2 + d(k)^2} \tag{1}$$

where $\lambda$ is the wavelength and $d(k)$ denotes the distance of the $k$th beam center from the center of the central beam and $d_0 \simeq 35786$ km.

*2) Rain Fading:* To model the rain attenuation effect we use the latest empirical model proposed in the ITU-R[2] Recommendation P.618 [19]. The distribution of the power gain $\xi$ in dB, $\xi_{\mathrm{dB}} = 20\log_{10}(\xi)$, is commonly modeled as a lognormal random variable, i.e., $\ln(\xi_{\mathrm{dB}}) \sim \mathcal{N}(\mu, \sigma)$, where $\mu$ and $\sigma$ depend on the location of the receiver, the frequency of operation, polarization and the elevation angle toward the satellite. The corresponding $K \times 1$ rain fading coefficients from all antenna feeds towards a single terminal antenna are given in the following vector

$$\tilde{\mathbf{h}} = \xi^{\frac{1}{2}} e^{-\mathrm{j}\phi \mathbf{1}_N} \tag{2}$$

where $\phi$ denotes a uniformly distributed phase. The phases from all antenna feeds are hard to differentiate and assumed to be identical. This is because we consider a LOS environment and the satellite antenna feed spacing is not large enough compared with the communication distance [20].

[2]International Telecommunications Union - Radiocommunications Sector.

Since rain attenuation is a slow fading process that exhibits spatial correlation over tens of kms, we assume that users undergo the same fading when located within the same beam, but independent fading among beams. In other words, we assume that each beam comprises a correlated area [13] [21].

*3) Beam Gain:* The link gain matrix defines the average signal to interference-plus-noise ratios (SINR) of the each user and it mainly depends on the satellite antenna beam pattern and the user position. Define one user's position based on the angle $\theta$ between the beam center and the receiver location with respect to the satellite and $\theta_{3\mathrm{dB}}$ is its 3-dB angle. Then the beam gain is approximated by [9]:

$$b\left(\theta,k\right)=\left(\frac{J_1\left(u\right)}{2u}+36\frac{J_3(u)}{u^3}\right)^2 \tag{3}$$

where $u=2.07123\sin\theta/\sin\theta_{3\mathrm{dB}}$ and $J_1$ and $J_3$ are the first-kind Bessel function of order 1 and 3. The coefficient $b_{\max}(k)$ represents the gain at the $k$th beam centre when FSL is taken into account as given by (1).

Collecting one user's beam gain coefficients from all transmit antenna into the $K\times 1$ vector $\mathbf{b}$, the overall channel for that user can be expressed as

$$\mathbf{h}=\sqrt{b_{\max}}\tilde{\mathbf{h}}\odot\mathbf{b}^{\frac{1}{2}}. \tag{4}$$

## *B. Signal Model*

Assume the data for user $k$ is $s_k$ with unit average power $\mathbb{E}[s_k^2]=1,\forall k$, where $k=1,2,\cdots,K$ is the index of users. The linear precoding vector $\mathbf{t}_k\triangleq\sqrt{p_k}\mathbf{w}_k$ is used to produce a weighted version of $s_k$ before transmission, where $p_k=\|\mathbf{t}_k\|^2$ is the transmit power and $\mathbf{w}_k=\frac{\mathbf{t}_k}{\|\mathbf{t}_k\|}$ is the normalized precoding vector. The purpose of introducing $p_k$ and $\mathbf{w}_k$ is that in some step of the proposed algorithm, they are separately optimized. The on board transmitted signal is the superposition of all users' signal, i.e., $\sum_{k=1}^{K}\mathbf{t}_k s_k$. The received signal at user $k$ can be written as

$$y_k=\mathbf{h}_k^{\dagger}\mathbf{t}_k s_k+\mathbf{h}_k^{\dagger}\sum_{j=1,j\neq k}^{K}\mathbf{t}_j s_j+n_k \tag{5}$$

where $n_k$ is the independent and identically distributed (i.i.d.) zero-mean Gaussian random noise. The received SINR for user $k$ is

$$\Gamma_k=\frac{|\mathbf{h}_k^{\dagger}\mathbf{t}_k|^2}{\sum_{j\neq k}|\mathbf{h}_k^{\dagger}\mathbf{t}_j|^2+N_0W}=\frac{p_k|\mathbf{h}_k^{\dagger}\mathbf{w}_k|^2}{\sum_{j\neq k}p_j|\mathbf{h}_k^{\dagger}\mathbf{w}_j|^2+N_0W} \tag{6}$$

where $N_0$ is the noise power density and $W$ is the total bandwidth. The achievable Shannon rate is

$$r_k=W\log(1+\Gamma_k),\forall k. \tag{7}$$

Denote the achievable rate vector as $\mathbf{r}\triangleq[\mathbf{r}_1,\cdots,\mathbf{r}_K]$.

We assume the satellite antenna beams are subject to both linear and nonlinear power constraints as described below.

## C. General Linear Power constraints

Suppose there are $L$ linear power constraints and the $l-$th constraint is expressed as

$$\sum_{k=1}^{K} \mathbf{t}_k^{\dagger}\mathbf{Q}_l\mathbf{t}_k \leq q_l, \forall l \tag{8}$$

where $q_l > 0$ is the power limit and $\mathbf{Q}_l \succeq \mathbf{0}$ is a shaping matrix which includes the following power constraints as special cases (we omit the constraint index $l$ for simplicity):

- Total beam power constraint: $\mathbf{Q} = \mathbf{I}, L = 1$;
- Per-beam (e.g., beam $k$) power constraint: $\mathbf{Q}_k = \mathbf{D}_k$ where $\mathbf{D}_k$ is a zero matrix except its $k$-th diagonal element being $1, L = K$;
- Flexible power constraints, e.g., $N$ beam antenna feeds in the set $\mathcal{Q} = \{k_1, \cdots, k_N\}$ have power sharing constraint: $\mathbf{Q}$ is a zero matrix except its diagonal elements with indices in $\mathcal{Q}$ being 1.

## D. General Nonlinear Power constraints

Although linear beam power constraints are widely adopted, nonlinear power constraints are also relevant. For example, the radio frequency (RF) output power of the high power amplifier (HPA) on board the satellite is a nonlinear function of the available direct current (DC) power, which is very costly. So if there is a limit on the total DC power, a constraint on nonlinear functions of beam powers is needed.

Suppose the nonlinear functions linking the beam output power $z_k$ to input power $x_k$ [3], is $z_k = g_k(x_k)$ where $g_k$ is a nonlinear function. Here we assume that $g_k$ is a continuous and increasing function and represents a one-to-one mapping between input and output, thus its inverse function $g^{-1}(\cdot)$ exists and $x_k = g_k^{-1}(z_k)$. We assume there are $J$ nonlinear power constraints with preset power limits $\{P_j\}$ on the beam antenna output power $\mathbf{t}_k^{\dagger}\mathbf{D}_k\mathbf{t}_k, \forall k$, i.e.,

$$\sum_{k=1}^{K} g_{k,j}^{-1}(\mathbf{t}_k^{\dagger}\mathbf{D}_k\mathbf{t}_k) \leq P_j, j = 1, \cdots, J \tag{9}$$

## E. Objective Functions

Suppose the required traffic demand for user $k$ is $F_k$. We consider a general objective function $f(\mathbf{r})$ of rates to optimize. We don't impose any special structure on $f(\mathbf{r})$ but we require that the function is continuous, which are satisfied by all commonly used performance metrics. Typical choices include

1) Throughput maximization: $\max \sum_{k=1}^{K} r_k$, which is a widely used performance metric in terrestrial communications;
2) Rate balancing: $\max \min_k \frac{r_k}{F_k}$, which is another way to meet the traffic demand while maximizing the worst user's rate to provide proportional fairness.
3) Rate matching: $\min \sum_{k=1}^{K} |F_k - r_k|^n$, where $n$ is a predefined order. This is especially useful for the design of satellite communications [8] to evaluate how well the traffic demand is met.

Different weights can also be introduced to indicate the priority or delay constraints for each user. We impose the additional constraint $r_k \leq F_k$, i.e., not to over satisfy any users' demand in order to save

[3] it may not necessarily be the input power to the amplifier and could be any power of interest

costly on board power and this has been included automatically in the rate matching problem. With this additional constraint, it is easily seen that throughput maximization is a special case of rate matching when $n = 1$ as

$$\sum_{k=1}^{K} |F_k - r_k| = \sum_{k=1}^{K} F_k - \sum_{k=1}^{K} r_k. \tag{10}$$

Note that different objectives in general result in different optimized rate vectors and the choice of the appropriate objective is up to the satellite operator.

### *F. Problem Formulation*

The problem of interest is to minimize a general objective function $f(\mathbf{r})$ of rates subjective to general linear and nonlinear power constraints by designing precoding vectors $\{\mathbf{t}_k, \forall k\}$. Mathematically, it is expressed as

$$\begin{aligned} \min_{\{\mathbf{t}_k\}} \quad & f(\mathbf{r}) \\ \text{s.t.} \quad & \sum_{k=1}^{K} \mathbf{t}_k^{\dagger} \mathbf{Q}_l \mathbf{t}_k \leq q_l, \forall l, \\ & \sum_{k=1}^{K} g_{k,j}^{-1}(\mathbf{t}_k^{\dagger} \mathbf{D}_k \mathbf{t}_k) \leq P_j, \forall j, \\ & r_k \leq F_k, \forall k. \end{aligned} \tag{11}$$

The main difficulty lies in the fact that the objective function $f(\mathbf{r})$ is non-convex in general, e.g., the throughput maximization problem is non-convex and has been studied in terrestrial communications [25]. The nonlinear (possibly non-convex) power constraints also introduces additional difficulty.

## III. Generic Precoding Design For Rate Enhancements

To tackle the difficulty of non-convex objective function and nonlinear power constraints, we adopt the strategy of alternating optimization [25], that is, we first optimize the precoding vectors $\{\mathbf{w}_k\}$ then update the power allocation $\{p_k\}$ with fixed precoding vectors.

### *A. Optimization of Precoding Vectors* $\{\mathbf{w}_k\}$

Optimizing the precoding vectors under the non-convex objective function and non-linear power constraints is too complex. Instead we propose to solve the following per-beam power minimization problem subject to minimum rate constraints, general linear power constraints (first kind) and per beam power constraints which are also linear (second kind):

$$\min_{\{\mathbf{t}_k, \gamma\}} \quad \gamma \tag{12}$$

$$\text{s.t.} \quad r_k \geq R_k, \forall k, \tag{13}$$

$$\sum_{k=1}^{K} \mathbf{t}_k^{\dagger} \mathbf{Q}_l \mathbf{t}_k \leq q_l, \forall l,$$

$$\sum_{k=1}^{K} \mathbf{t}_k^{\dagger} \mathbf{D}_j \mathbf{t}_k \leq \gamma \tilde{P}_j, j = 1, \cdots, K, \tag{14}$$

where $\tilde{P}_j$ is a known parameter related to the $j$-th beam's power and $\gamma$ is an auxiliary variable. The above problem aims to reduce each beam's transmit power while satisfying the rate constraints $\{R_k\}$, which is a predefined minimum rate requirement. The reason of such a formulation will become clear as our discussion goes on.

Note that the above optimization problem is convex since the objective function is convex, the two kinds of linear power constraints are convex and the individual rate constraints (13), although originally non-convex, can be made convex using the transformation in [26] and rewritten as

$$\sqrt{(2^{\frac{R_k}{W}} - 1)\left(\sum_{j\neq k} |\mathbf{t}_j^\dagger \mathbf{h}_k|^2 + WN_0\right)} \leq \mathbf{h}_k^\dagger \mathbf{t}_k \tag{15}$$

and as a result problem (12) is recognized as a second-order cone programming (SOCP) problem which is convex and readily solved using standard techniques [27].

An interesting fact about (12) is that at the optimum, the rate constraints $R_k = r_k, \forall k$, which aligns automatically with the last constraint in the original problem (11).

Once we get the optimal $\mathbf{t}_k$, the normalized precoding vector can be obtained as $\mathbf{w}_k = \frac{\mathbf{t}_k}{\|\mathbf{t}_k\|}$. The power is also optimized in (12) via $\mathbf{t}_k$, although it does not take into account the general objective function $f(\mathbf{r})$, it is still very useful in developing the algorithm and for this purpose the beam power is stored as $\hat{p}_k = \|\mathbf{w}_k\|^2, \forall k$ for later use.

Notice that the problem (12) is also useful to detect whether the traffic demand can be satisfied or not, i.e., replace $R_k$ with $F_k$ and solve (12). If it is feasible, then the optimization is done as the traffic demand is perfectly met. In the remainder of this paper, we always assume that the traffic demand is high enough such that it is not possible to satisfy all users' rate constraints simultaneously.

### *B. Power Optimization*

With fixed precoding vectors $\{\mathbf{w}_k\}$, the general power optimization problem can be modified from (11) and formulated as

$$\begin{aligned}
\min_{\{p_k\}} \quad & f(\mathbf{r}) && (16)\\
\text{s.t.} \quad & \sum_{k=1}^{K} p_k \mathbf{w}_k^\dagger \mathbf{Q}_l \mathbf{w}_k \leq q_l, \forall l,\\
& \sum_{k=1}^{K} g_{k,j}^{-1}(\mathbf{t}_k^\dagger \mathbf{D}_j \mathbf{t}_k) \leq P_j, \forall j,\\
& r_k = W \log_2\left(1 + \frac{p_k |\mathbf{h}_k^\dagger \mathbf{w}_k|^2}{\sum_{j\neq k} p_j |\mathbf{h}_k^\dagger \mathbf{w}_j|^2 + N_0 W}\right) \leq F_k, \forall k.
\end{aligned}$$

It is seen that both the linear power constraints and the maximum rate constraints are in linear form about power $\{p_k\}$ which are easy to handle. However, due to the possibly non-convex objective function $f(\mathbf{r})$ and nonlinear non-convex power constraints, (16) is non-convex in general and its globally optimal solution is difficult to find.

Here we propose to find its local optimal solution by using any gradient-based numerical algorithm, such as the steepest descent algorithm, sequential quadratic programming method and trust region optimization [28]. Although the globally optimal power solution is not guaranteed to (16), it will be seen later that by properly choosing the initial point, a convergent iteration algorithm can be constructed.

### *C. The Proposed Generic Iterative Algorithm and Proof of Convergence*

With the above strategies to optimize precoding vectors and power allocation alternatingly, we propose a generic iterative algorithm below to jointly optimize the precoding vectors and power allocation.

**Proposed Generic MISO Algorithm**

Step 1: Initialize $\{\mathbf{w}_k\}$ and $\{p_k\}$ such that both the linear power constraints and maximum rate constraints are satisfied in (11). Towards this, a convenient choice is that $\{p_k\}$ are chosen to be very small values.

Step 2: For each $k$, evaluate the achievable rate for user $k$ using (6-7) and store it as $R_k$.

Step 3: With the above $\{R_k\}$ as inputs, solve the optimization problem (12) for $\mathbf{t}_k$, then obtain the precoding vector $\mathbf{w}_k = \frac{\mathbf{t}_k}{\|\mathbf{t}_k\|}$, the power solution $\hat{p}_k = \|\mathbf{t}_k\|^2$, and $\tilde{P}_k = \mathbf{t}_k^\dagger \mathbf{D}_j \mathbf{t}_k, \forall k$.

Step 4: With fixed $\{\mathbf{w}_k\}$, solve the optimization problem (11) for power allocation $\{p_k\}$ using a gradient-based algorithm with $\{\hat{p}_k\}$ being the initial power solution.

Step 5: Go back to Step 2 until it converges.

Although there are many possibilities to construct iterative algorithms based on alternating optimization, the convergence is not guaranteed. Next we will prove that the proposed generic MISO algorithm indeed converges.

*Theorem 1:* The proposed Algorithm 1 always converges.

*Proof:*

We denote the value of objective function $f(\mathbf{r})$ at Step 2,3,4 of the $n$th iteration as $f_2^{(n)}, f_3^{(n)}$ and $f_4^{(n)}$, respectively.

Due to the small initial power $\{p_k\}$, $R_k \le F_k, \forall k$ always holds true. From Step 2 to Step 3, it is easy to see that the problem (12) at Step 3 is always feasible and $R_k$ is satisfied with equality. Therefore we have

$$f_2^{(n)} = f_3^{(n)}. \tag{17}$$

The improvement at Step 3 is that less or at most equal power is used for each beam as that at Step 2, which in turn satisfies all nonlinear beam power constraints (due to their monotonicity) and creates room for further power optimization. This is realized by introducing the per beam power constraints (14).

At Step 4, the power is further optimized locally and the power solution $\{\hat{p}_k\}$ at Step 3 is taken as an initial point such that the optimized power $\{p_k^*\}$ should be no worse than $\{\hat{p}_k\}$ which is obtained based on another criterion of minimizing beam power rather than $f(\mathbf{r})$, in other words,

$$f_4^{(n)} \le f_3^{(n)}. \tag{18}$$

Combining (17) and (18), we have

$$f_4^{(n)} \le f_3^{(n)} = f_2^{(n)} = f_4^{(n-1)}, \tag{19}$$

which implies that the value of the objective function is monotonically decreasing from one iteration to the next and as it is lower bounded by zero, the algorithm always converges. This completes the proof. ■

We have the following remarks on solving problem (12) at Step 3 of the proposed algorithm, which is a key step in the algorithm.

*Remark 1:* We adopt the maximum beam power as the objective function in (12) and therefore the outcome of the algorithm is that we minimize the general objective function $f(\mathbf{r})$ with minimum beam power. This choice of objective is general enough while for a specific problem, it is not unique, e.g., if there are only linear power constraints, the objective function at Step 3 can be replaced by the maximum user power or total transmit power which indicates their respective importance. With these changes of objective function at Step 3, the overall convergence is still guaranteed which can be seen from the proof of Theorem 1.

*Remark 2:* Although (12) can be solved using standard numerical algorithm, there are less complex algorithms by making use of its analytical structure. We propose a more efficient algorithm in the Appendix to solve (12) based on its dual problem and subgradient method inspired by [29]. The proposed algorithm in the Appendix can be modified to deal with different objective functions.

## IV. An Extension To DPC and A Special Case For Fairness Maximization

### A. Application to DPC

DPC is known to be the sum capacity-achieving technique in multiuser MIMO downlink. Here we will show that given a decoding order, the proposed MISO algorithm can be applied to the precoding design in DPC straightforwardly.

Let us assume that $\pi_0 = \{1, 2, ..., K\}$ is a trivial user encoding order. Then the received SINR at user $k$ is

$$\Gamma_k = \frac{|\mathbf{h}_k^\dagger \mathbf{t}_k|^2}{\sum_{j>k} |\mathbf{h}_k^\dagger \mathbf{t}_j|^2 + N_0 W} \tag{20}$$

which possesses similar structure as (6) and thus the proposed algorithm can be easily modified to find the precoding vectors for DPC.

Since determining the optimal encoding order is computationally expensive, we investigate the performance of some intuitive heuristics. In general, users encoded first receive a large amount of interference which gradually decreases as the serial encoding proceeds. In this respect, users with good channel conditions should be encoded first because they can afford being interfered. Similarly, users with high rate requirements should be decoded last because they have to avoid interference to achieve the requested rates. As a result, the users are ordered based on the increasing order of the metric: $\left\{\frac{F_k}{\log_2(1+\|\mathbf{h}_k\|^2)}\right\}$.

### B. Special Case: Optimal and Efficient Solution to Fairness Maximization with Convex Power Constraints

As mentioned before, in general the objective function $f(\mathbf{r})$ is non-convex which makes the whole problem difficult even with linear power constraints only. There has been a special case well studied in terrestrial communications [30] where the efficient and globally optimal solution exists: to maximize the worst user's weighted rate in order to achieve fairness subject to the total power constraint. Here we extend it to the nonlinear but convex power constraints. Mathematically, the problem is formulated as

$$\begin{aligned} \max_{\{\mathbf{t}_k\}} \quad & \min_k \frac{r_k}{F_k} \\ \text{s.t.} \quad & \sum_{k=1}^{K} \mathbf{t}_k^\dagger \mathbf{Q}_l \mathbf{t}_k \le q_l, \forall l, \\ & \sum_{k=1}^{K} g_{k,j}^{-1}(\mathbf{t}_k^\dagger \mathbf{D}_j \mathbf{t}_k) \le P_j, \forall j, \\ & r_k \le F_k, \forall k, \end{aligned} \tag{21}$$

where we assume $g_{k,j}^{-1}(\cdot)$ is a convex function. Next we propose an algorithm within the framework of optimization to achieve the optimum. We first introduce an auxiliary variable $\gamma$ and reformulate it into

$$\begin{aligned} \max_{\{\mathbf{t}_k\},\gamma\le 1} \quad & \gamma \\ \text{s.t.} \quad & \mathbf{t}_k^\dagger \mathbf{h}_k \ge \sqrt{\left(2^{\frac{\gamma F_k}{W}} - 1\right)\left(\sum_{j\ne k} |\mathbf{t}_j^\dagger \mathbf{h}_k|^2 + W N_0\right)},\ \forall k, \\ & \sum_{k=1}^{K} \mathbf{t}_k^\dagger \mathbf{Q}_l \mathbf{t}_k \le q_l, \forall l \\ & \sum_{k=1}^{K} g_{k,j}^{-1}(\mathbf{t}_k^\dagger \mathbf{D}_j \mathbf{t}_k) \le P_j, \forall j. \end{aligned} \tag{22}$$

Now (22) is a convex problem except for the variable $\gamma$. To solve it, we use the bi-section search approach [27] to find the optimum solution. To be specific, at each search iteration for a given $\tilde{\gamma} \le 1$ (this guarantees $r_k \le F_k, \forall k$), we check whether all constraints in (22) are feasible or not. If it is feasible, then increase the value of $\tilde{\gamma}$ and decrease it otherwise.

## V. Multiple Antennas at Receive Terminals

In this section, we assume the user terminals have multiple antennas and the system turns out to be a multiuser MIMO downlink system. Specifically we study the effects of co-polarization and dual polarization. Dual polarization has been proposed as a viable solution for employing single-user MIMO techniques in mobile satellite systems [31], but its performance in fixed multibeam satellite systems is not well understood. In this section, we investigate the effect of dual-polarization antennas with linear precoding in the FL of fixed multibeam satellite systems.

Assume that each terminal is equipped with multiple antennas. More specifically, the following four cases are considered: 1) two co-polarization antennas with receive beamforming (RBF), 2) two cross-polarization antennas with RBF, 3) two co-polarization antennas with antenna selection, 4) two cross-polarization antennas with polarization selection. In general,RBF is optimal but antenna selection is less expensive in computation and implementation. In terms of polarization, cross-polarization antennas achieve lower receive SNR because of the power imbalance but they have better de-correlation properties due to propagation characteristics. As a result it is not straightforward to judge which scheme is better and the main aim is to choose the best technique based on realistic channel parameters.

### A. Channel Modeling

*1) Two co-polarization antennas with receive beamforming:* Assume each of $K$ terminals has two receive antennas. The channel for each terminal can be denoted as $\mathbf{H}_k$ with dimensions $2 \times K$. A main assumption of the considered channel model is full transmit-side correlation due to the lack of scatterers close to the satellite. For the co-polarization case, the two receive antennas are also assumed to be fully correlated and thus the channel for the $k$th user can be written as

$$\mathbf{H}_k = \mathbf{1}_2 \otimes \mathbf{h}_k, \tag{23}$$

where $\mathbf{h}_k$ is defined in (4). The beam gain coefficients for each user vary across the beams due to the beam pattern, while rain fading is identical due to identical illumination angle.

*2) Two cross-polarization antennas with receive beamforming:* Assume that both satellite feeds and terminal antennas are dual-polarized. For the sake of simplicity, a Kronecker correlation model between polarizations is assumed with both transmit and receive correlation. In this direction, the channel for the $k$th user can be written as the $2 \times 2N$ matrix:

$$\bar{\mathbf{H}}_k = \sqrt{b_{\max}(k)} \left( \mathbb{I}_2 \otimes \mathbf{b}_k^{\frac{1}{2}} \right) \odot \left( \mathbf{1}_N^T \otimes \left( \mathbf{R}^{\frac{1}{2}} \tilde{\mathbf{H}}_k \mathbf{R}^{\frac{1}{2}} \right) \right) \odot \left( \mathbf{1}_N^T \otimes \mathbf{P} \right), \tag{24}$$

where $\mathbf{P}$ is a $2 \times 2$ matrix modelling the power imbalance between polarizations, $\tilde{\mathbf{H}}_k$ includes the rain fading coefficients for co- and cross-polarization subchannels and $\mathbf{R}$ is the correlation matrix between polarizations[4]. More specifically, the matrix $\tilde{\mathbf{H}}_k$ can be written as:

$$\tilde{\mathbf{H}}_k = \begin{bmatrix} \xi_{00}^{\frac{1}{2}} & \xi_{01}^{\frac{1}{2}} \\ \xi_{10}^{\frac{1}{2}} & \xi_{11}^{\frac{1}{2}} \end{bmatrix} e^{-\mathrm{j}\phi}, \tag{25}$$

where $\xi_{ij}, \forall i,j = \{0,1\}$ are assumed to be independent rain fading coefficients as modelled in Section II.A. $\phi$ is a uniformly distributed phase. Furthermore, if $\alpha$ is the power imbalance factor and $\rho$ is the correlation factor, $\mathbf{P}$ and $\mathbf{R}$ can be modelled as:

$$\mathbf{P} = \begin{bmatrix} 1 & \alpha \\ \alpha & 1 \end{bmatrix} \tag{26}$$

$$\mathbf{R} = \begin{bmatrix} 1 & \rho \\ \rho & 1 \end{bmatrix} \tag{27}$$

It should be noted that the $\alpha$ factor is related to Cross Polarization Discrimination (XPD) as follows: $\alpha = (\sqrt{\mathrm{XPD}})^{-1}$.

[4]It should be noted that due to the statistics of the rain fading coefficients (non-zero mean), the channel power of $\bar{\mathbf{H}}_k$ will be higher than the one of $\mathbf{H}_k$. In order to ensure a fair comparison amongst all scenarios, the channel matrices are normalized so that $\mathrm{trace}(\bar{\mathbf{H}}_k \bar{\mathbf{H}}_k^\dagger) = \mathrm{trace}(\mathbf{H}_k \mathbf{H}_k^\dagger)$.

### B. Receive Strategy I: Antenna Selection

*1) Two co-polarization antennas with antenna selection:* Since in the co-polarization case the two antennas are considered fully correlated, this scenario degrades to the single-antenna terminal already studied in the previous sections.

*2) Two cross-polarization antennas with polarization selection:* Building on scenario 2 we discern two cases. In the first case, each satellite antenna has single polarization. The corresponding $2 \times N$ channel matrix can be easily derived from (24) by picking alternating polarizations. In the second case, we also select the best polarization at the terminal using the criterion

$$\arg\max_{n} \|\mathbf{H}_k(n,:)\|, \forall k \tag{28}$$

which results in a $1 \times N$ channel matrix.

Both of the above cases result in equivalent multiuser MISO channels, therefore the proposed generic MISO algorithm could be applied to find the optimized precoding vectors.

### C. Receive Strategy II: RBF and The Proposed Generic MIMO Algorithm

When there are multiple receive antennas at the terminal, assuming single-user decoding, the optimal strategy is to use RBF vector $\{\mathbf{u}_k\}$ that matches the channels and the precoders. The received SINR for user $k$ can be expressed as

$$\Gamma_k = \frac{|\mathbf{u}_k^\dagger \mathbf{H}_k \mathbf{t}_k|^2}{\sum_{j \neq k} |\mathbf{u}_k^\dagger \mathbf{H}_k \mathbf{t}_j|^2 + N_0 W} \tag{29}$$

and for fixed precoding vectors $\mathbf{t}$, the optimal RBF vector to maximize the received SINR is given by

$$\mathbf{u}_k = \left( \sum_{j=1}^{K} \mathbf{H}_k \mathbf{t}_j \mathbf{t}_j^\dagger \mathbf{H}_k^\dagger + N_0 W \mathbf{I} \right)^{-1} \mathbf{H}_k \mathbf{t}_k. \tag{30}$$

With the above result, we propose the following generic MIMO algorithm based on the proposed generic MISO algorithm.

**Proposed Generic MIMO Algorithm**

Step 1: Initialize $\{\mathbf{w}_k, \mathbf{u}_k, p_k\}$ such that both the linear power constraints and maximum rate constraints are satisfied in (11). To guarantee that, normally $\{p_k\}$ need to be very small values.

Step 2: For each $k$, evaluate the achievable rate for user $k$ using (29) and store it as $R_k$.

Step 3: With the above $\{R_k\}$ as inputs and given fixed $\{\mathbf{u}_k\}$, solve the optimization problem similar to (12) with substitution $\mathbf{h}_k^\dagger \triangleq \mathbf{u}_k^\dagger \mathbf{H}_k, \forall k$ for $\{\mathbf{t}_k\}$, then obtain the precoding vector $\mathbf{w}_k = \frac{\mathbf{t}_k}{\|\mathbf{t}_k\|}$, the power solution $\tilde{p}_k = \|\mathbf{t}_k\|^2$ and $\tilde{P}_k = \mathbf{t}_k^\dagger \mathbf{D}_j \mathbf{t}_k, \forall k$.

Step 4: Update the RBF vector $\{\tilde{\mathbf{u}}_k\}$ using (30).

Step 5: Evaluate the achievable rate for user $k$ using (29) and store it as $\tilde{R}_k, \forall k$.

Step 6: If $\tilde{R}_k < F_k$, then update $\mathbf{u}_k = \tilde{\mathbf{u}}_k, \forall k$.

Step 7: With fixed $\{\mathbf{w}_k, \mathbf{u}_k\}$, solve the optimization problem similar to (11) with substitution $\mathbf{h}_k^\dagger \triangleq \mathbf{u}_k^\dagger \mathbf{H}_k, \forall k$ for power allocation $\{p_k\}$ using a gradient-based algorithm with $\{\tilde{p}_k\}$ being the initial power solution.

Step 8: Go back to Step 2 until it converges.

The difference between the proposed MIMO algorithm and the proposed MISO algorithm is due to Step 4-6, which update the RBF vectors and ensures the rate requirements are not over satisfied. The proposed generic MIMO Algorithm can be proved to always converge using the same argument in the proof of Theorem 1 and thus the proof is omitted here.

## VI. Simulation Results

Computer simulations are conducted to evaluate the performance of the proposed algorithms. For multibeam joint precoding, we consider a fixed satellite system as described in Section II-A with detailed parameters listed in Table I. The rain fading corresponds to temperate central European climate. We assume there are K=7 on board antenna feeds serving 7 beams on the ground. Within each beam, there are 4 fixed user terminals and they are served in a TDM manner. Achievable rates for all users will be shown as the performance metric for different schemes and different objectives. Unless otherwise specified, we choose the $l_2$ norm minimization, $f(\mathbf{r}) = \sum_{k=1}^{K} |F_k - r_k|^2$, as the objective function. Traffic demand is assumed to be asymmetric and uniformly distributed with the mean vector listed in Table I. For linear power constraints, as listed in Table I, we assume each satellite RF saturation power is 80 dBW, which is the individual beam power constraint.

TABLE I
SATELLITE SCENARIO PARAMETERS

| Parameter | Value |
|---|---|
| Orbit | GEO |
| Frequency band | 20 GHz |
| Number of beams | $K = 7$ |
| Beam diameter | $D = 250$ km |
| 3dB angle | $\theta_{3\text{dB}} = 0.4^o$ |
| Rain fading mean | $\mu = -2.6$ |
| Rain fading variance | $\sigma = 1.63$ |
| Polarization | Single/Dual |
| Max antenna Tx gain | 52dBi |
| TWTA RF power @ saturation | 80W |
| User terminal maximum antenna gain | 41.7dBi |
| FL free space loss | 210dB |
| User link bandwidth | W=500MHz |
| Clear sky receiver temperature | $207^o$K |
| Mean of uniformly distributed traffic demand | [4 0.8 0.8 0.8 2 2 2] Gbps |
| Frequency reuse factor for the conventional scheme | 4 |

For the MISO case where the user terminal has only one receive antenna and single polarization, the proposed generic MISO optimization algorithm will be compared with the following schemes:

1) Conventional schemes with single-beam processing, the same bandwidth $W$ and frequency reuse factor 4. The achievable rate for user $k$ is

$$r_k^b = \frac{W}{4} \log_2 \left(1 + \frac{4p_k |h_{k,k}|^2}{W N_0}\right), \tag{31}$$

where $p_k$ is the transmit power for beam $k$ and co-channel beam interference is ignored;

2) ZF precoding: collect all users's channels into a $K \times K$ matrix $\tilde{\mathbf{H}} = [\mathbf{h}_1^\dagger; \cdots; \mathbf{h}_K^\dagger]$, then precoding vector $w_k$ is taken from the normalized $k$-th column of $(\tilde{\mathbf{H}}^\dagger \tilde{\mathbf{H}})^{-1} \tilde{\mathbf{H}}^\dagger$
3) R-ZF precoding: $w_k$ is taken from the normalized $k$-th column of $(\tilde{\mathbf{H}}^\dagger \tilde{\mathbf{H}} + a\mathbf{I})^{-1} \tilde{\mathbf{H}}^\dagger$ where $a = \frac{N_0 W}{P_0}$ [33] and $P_0 = 80$ W is the maximum beam power constraint;
4) DPC with the encoding order determined by $\left\{ \frac{F_k}{\log_2(1+\|\mathbf{h}_k\|^2)} \right\}$ and the nonlinear precoders are optimized using the proposed algorithms to provide a performance upper bound.

Performance results are depicted for all 7 beams whose indices are shown on x-axis. Fig. 2 shows individual users' rates for different MISO schemes to match the traffic demand. The throughput and the average value of the $l_2$ norm cost function ($n = 2$ in the rate matching function in Section II.E) are shown in the legend. First it is verified that all multibeam schemes outperform the conventional single-beam processing. The simple ZF precoding can achieve 67% more throughput than the conventional scheme. R-ZF performs slightly better than ZF by considering the noise effects. Compared with ZF precoders, it is observed that the proposed generic optimization scheme achieves slightly higher throughput but matches the traffic much better (40% lower $l_2$ norm). It is also seen that DPC only has very marginal performance gain (around 4% in terms of throughput and 6% for $l_2$ norm) over the proposed linear precoding. We also study the impact of flexible total power constraint of $7 \times 80 = 560$ W for the proposed scheme and DPC. It is seen that higher rates are achieved compared with the individual power constraints, which is expected. The proposed scheme with the total power constraint even outperforms DPC with individual power constraints, which implies that flexible power constraints are more important than the complicated non-linear signal processing. It is again observed that the proposed linear precoding performs nearly as well as DPC. This indicates that linear processing is adequate for multibeam satellite.

Fig. 3 shows the average actual power used by each beam for different transmission schemes. The total power is also shown in the legend. First it is noted that all schemes have a tendency to allocate power adaptively to traffic demand. The conventional scheme always uses the highest amount of power due to the lack of multibeam cooperation. ZF solution uses only about 60% total power of that used by the conventional scheme which is a huge saving of satellite power. The proposed scheme consumes less power but more adaptively than ZF to meet the traffic demand within the per beam power limit. To get a clearer picture of achieved rate and used power, Fig. 4 shows the rate efficiency normalized by the actual power used by each beam. It is confirmed that the conventional scheme has a very low rate efficiency due the low rate achieved and high power consumed. ZF solution has a high efficiency for those beams with low traffic demand due to the low power used.

Fig. 5 illustrates the impacts of choosing rate balancing, $l_2$ norm minimization, throughput maximization as objective functions. It is seen that although the objective of rate balancing guarantees fairness among users, it degrades too much the throughput performance. The proposed schemes with sum rate maximization and $l_2$ minimization as objectives greatly outperform rate balancing and achieve 10% and 8% higher throughput, respectively. Compared with DPC precoding which is optimal for throughput maximization, the performance degradation of the proposed linear precoding is almost negligible.

To evaluate the performance of the MIMO case when user terminals have either co-polarization or dual-polarization antennas, the proposed generic MISO and MIMO optimization algorithms are applied

to the following cases:

| Case | Feeds | Terminals | Technique |
|---|---|---|---|
| 1 | Single-polarization | 2 co-polarization antennas, $\rho = 1$ | RBF |
| 2 | Dual-polarization | 2 dual-polarization antennas, $\rho = 0.9$ | RBF |
| 3 | Alternating single-polarization[5] | 2 dual-polarization antennas, $\rho = 0.9$ | RBF |
| 4 | Single-polarization selection | 2 dual-polarization antennas, $\rho = 0.9$ | Best polarization selection |

The results are shown in Fig. 6. The throughput and the averaged value of the $l_2$ norm cost function are shown in the legend. It is observed that with co-polarization receive antennas, the $l_2$ norm is significantly reduced compared with the MISO case, which indicates better traffic matching. The scheme with dual-polarization at satellite antenna feeds and terminals employing RBF achieves the best performance at the cost of increased hardware complexity. The performance of the scheme with alternating single-polarization satellite and dual-polarization terminals employing RBF is comparable to that with co-polarization receive antennas, while polarization selection further degrades the performance. The performance of all dual polarization techniques is greatly degraded by the correlation effects. The co-polarization receive antennas at the terminal seem to be a more promising solution considering both performance and hardware complexity.

## VII. Conclusions

This paper has studied the optimization of linear precoding for multibeam satellites with general linear and nonlinear power constraints. A generic iterative algorithm has been proposed to handle any objective function of individual rates and the convergence is proved. The proposed algorithm has been extended to design DPC precoding with fixed encoding order and to the case when user terminals are equipped with two co-polarization or dual-polarization receive antennas. Simulation results have shown substantial performance gain compared with conventional single beam based processing and existing precoders. The optimized linear precoding is also shown to be as effective as DPC precoding, which confirms its practical value. The impact of co/dual-polarization has also been demonstrated.

Future work includes the study of of promising non-linear precoding design like THP in a DVB-S2 system. In addition, in this paper, the feeder link and CSI at the GW are assumed perfect, which is not very practical and will be investigated in the future.

[5]Each feed is single polarization but we alternate between co- and cross-polarization as we enumerate the feeds.

# APPENDIX
## EFFICIENT ALGORITHM TO SOLVE (12)

The dual problem of (12) is

$$\max_{\boldsymbol{\alpha}\geq 0,\boldsymbol{\lambda}\geq 0,\boldsymbol{\mu}\geq 0} \quad WN_0 \sum_{k=1}^{K} \alpha_k - \sum_{l=1}^{L} \lambda_l q_l \tag{32}$$
$$\text{s.t.} \quad \sum_{j=1}^{K} \mu_j \mathbf{D}_j + \sum_{l=1}^{L} \lambda_l \mathbf{Q}_l + \sum_{\tilde{k}=1}^{K} \alpha_k \mathbf{h}_{\tilde{k}} \mathbf{h}_{\tilde{k}}^\dagger \succeq \left(1 + \frac{1}{2^{\frac{R_k}{W}} - 1}\right) \mathbf{h}_k \mathbf{h}_k^\dagger, \forall k,$$
$$\sum_{j=1}^{K} \mu_j P_j \leq \sum_{j=1}^{K} P_j.$$

The strategy is to first fix $\boldsymbol{\lambda}$ and $\boldsymbol{\mu}$ and solve (32) for $\boldsymbol{\alpha}$ and find corresponding primary solution $\{\mathbf{t}_k\}$ then use it to update $\boldsymbol{\lambda}$ and $\boldsymbol{\mu}$.

Given fixed $\boldsymbol{\lambda}$, (32) reduces to

$$\max_{\boldsymbol{\alpha}\geq 0} \quad WN_0 \sum_{k=1}^{K} \alpha_k \tag{33}$$
$$\text{s.t.} \quad \sum_{j=1}^{K} \mu_j \mathbf{D}_j + \sum_{l=1}^{L} \lambda_l \mathbf{Q}_l + \sum_{\tilde{k}=1}^{K} \alpha_k \mathbf{h}_{\tilde{k}} \mathbf{h}_{\tilde{k}}^\dagger \succeq \left(1 + \frac{1}{2^{\frac{R_k}{W}} - 1}\right) \mathbf{h}_k \mathbf{h}_k^\dagger, \forall k.$$

whose optimal solution has the interpretation of virtual uplink power and is found by performing the following iterative update [32]:

$$\alpha_k = \frac{1}{\left(1 + \frac{1}{2^{\frac{R_k}{W}} - 1}\right) \mathbf{h}_k^\dagger \left(\mathbf{I} + \sum_{l=1}^{L} \lambda_l \mathbf{Q}_l + \sum_{k=1}^{K} \alpha_k \mathbf{h}_k \mathbf{h}_k^\dagger\right)^{-1} \mathbf{h}_k}. \tag{34}$$

Then the optimal precoding must be in the direction of

$$\tilde{\mathbf{t}}_k = \left(\sum_{j=1}^{K} \mu_j \mathbf{D}_j + \sum_{l=1}^{L} \lambda_l \mathbf{Q}_l + \sum_{k=1}^{K} \alpha_k \mathbf{h}_k \mathbf{h}_k^\dagger\right)^{-1} \mathbf{h}_k \tag{35}$$

and the downlink power can be obtained by using the fact that the rate constraints are satisfied with equalities in (12).

Summarizing the above results, we propose an efficient algorithm using similar subgradient method in [29] as follows. For more details, please refer to [29].

**Proposed Efficient Algorithm to Solve (12)**

Step 1: Initialize $\boldsymbol{\lambda} \geq 0, \boldsymbol{\mu} \geq 0$.

Step 2: Given $\boldsymbol{\lambda}, \boldsymbol{\mu}$, find the optimal solution to (33) by using fixed point update (34) iteratively.

Step 3: Update $\tilde{\mathbf{t}}_k$ using (35), $\forall k$.

Step 4: Find downlink power using

$$\boldsymbol{\delta} = \mathbf{G}^{-1}\mathbf{1}WN_0 \tag{36}$$

where $\mathbf{G}_{kk} = \frac{1}{2^{\frac{R_k}{W}}-1}|\tilde{\mathbf{t}}_k^{\dagger}\mathbf{h}_k|^2, \mathbf{G}_{kj} = -|\tilde{\mathbf{t}}_j^{\dagger}\mathbf{h}_k|^2, \forall j \neq k$.

Step 5: Update downlink precoding vectors $\mathbf{t}_k = \sqrt{\delta_k}\tilde{\mathbf{t}}_k, \forall k$.

Step 6: Update $\lambda_l = \max(0, \lambda_l + \eta_l \Delta\lambda_l), \forall l$ where $\eta_l$ is the step size. $\Delta\lambda_l$ is the subgradient and one possible choice is $\Delta\lambda_l = \sum_{k=1}^{K} \mathbf{t}_k^{\dagger}\mathbf{Q}_l\mathbf{t}_k - q_l, \forall l$.

Step 7: Update $\mu_j = \max(0, \mu_j + \rho_j \Delta\mu_j), \forall l$ where $\rho_j$ is the step size. $\Delta\mu_j$ is the subgradient and one possible choice is $\Delta\mu_j = \sum_{k=1}^{K} \mathbf{t}_k^{\dagger}\mathbf{D}_l\mathbf{t}_k, j = 1, \cdots, K$.

Step 8: Find valid $\boldsymbol{\mu}$ using the projection method into the feasible set $\{\boldsymbol{\mu} : \sum_{j=1}^{K} \mu_j P_j \leq \sum_{j=1}^{K} P_j\}$.

Step 9: Go back to Step 2 until it converges.

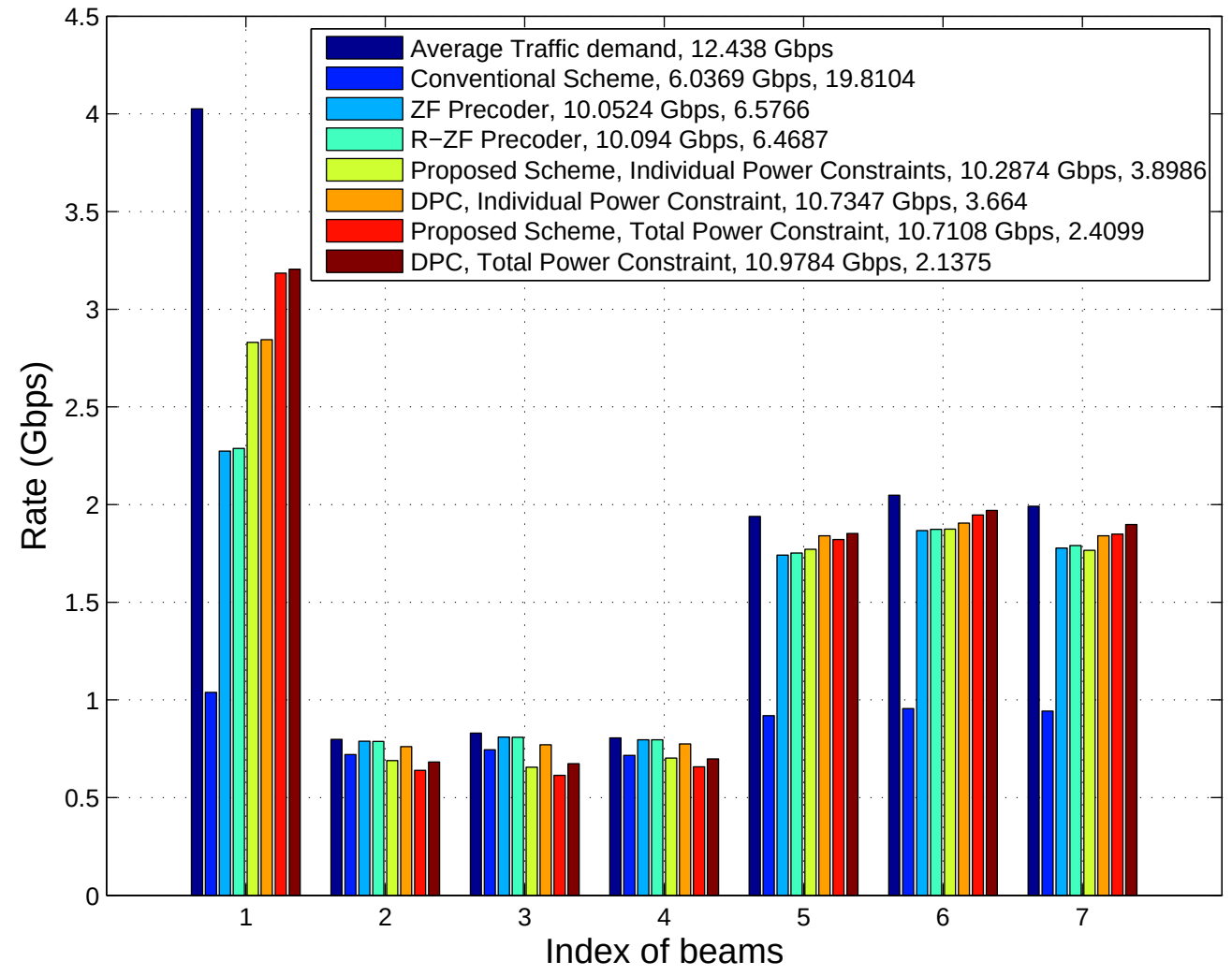

Fig. 2. Comparison of rates for different schemes with the objective of $l_2$ norm minimization. In the legend, the name of each scheme is followed by the throughput and the average value of the $l_2$ norm.

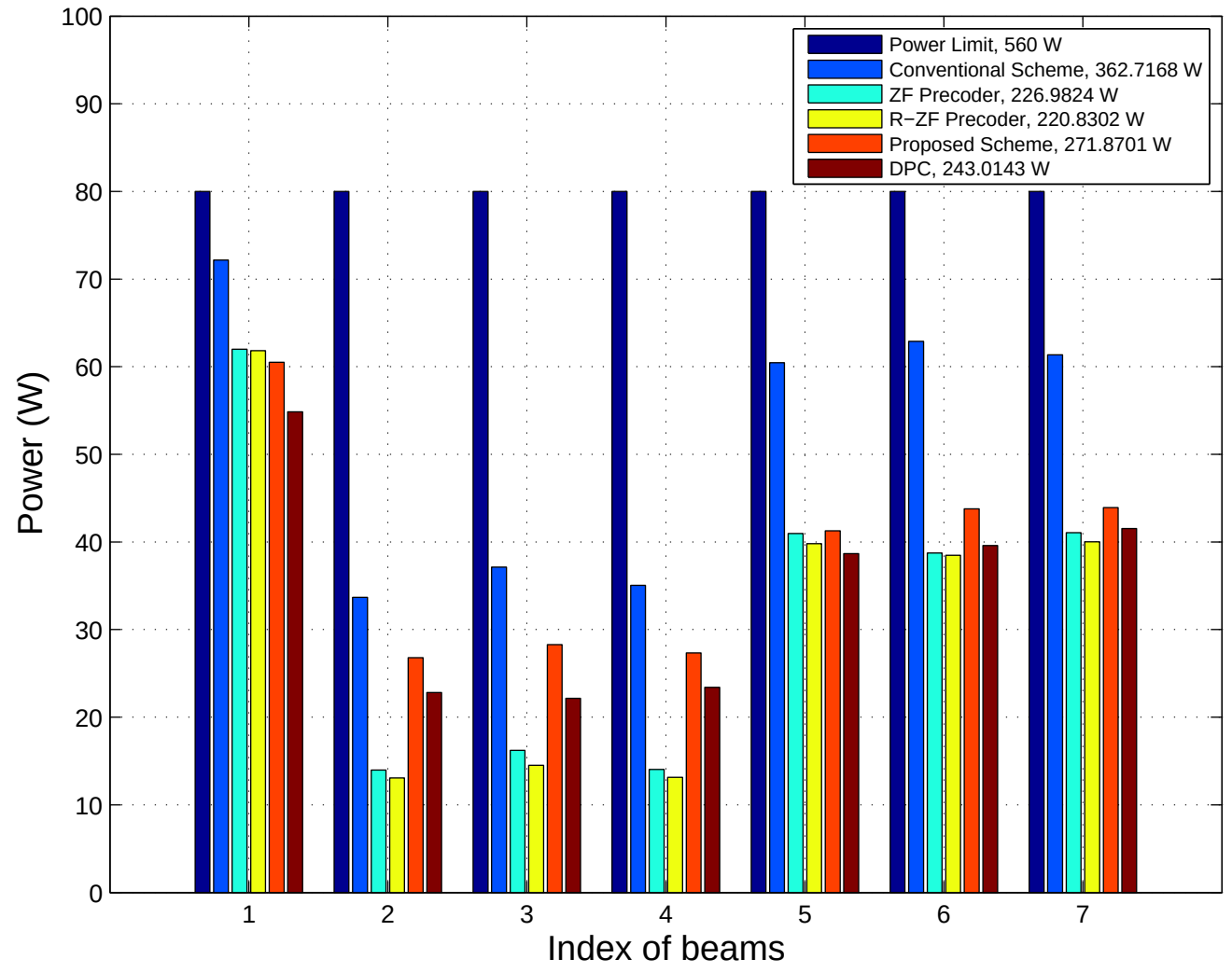

Fig. 3. Comparison of beam powers for different schemes. In the legend, the name of each scheme is followed by the total beam power.

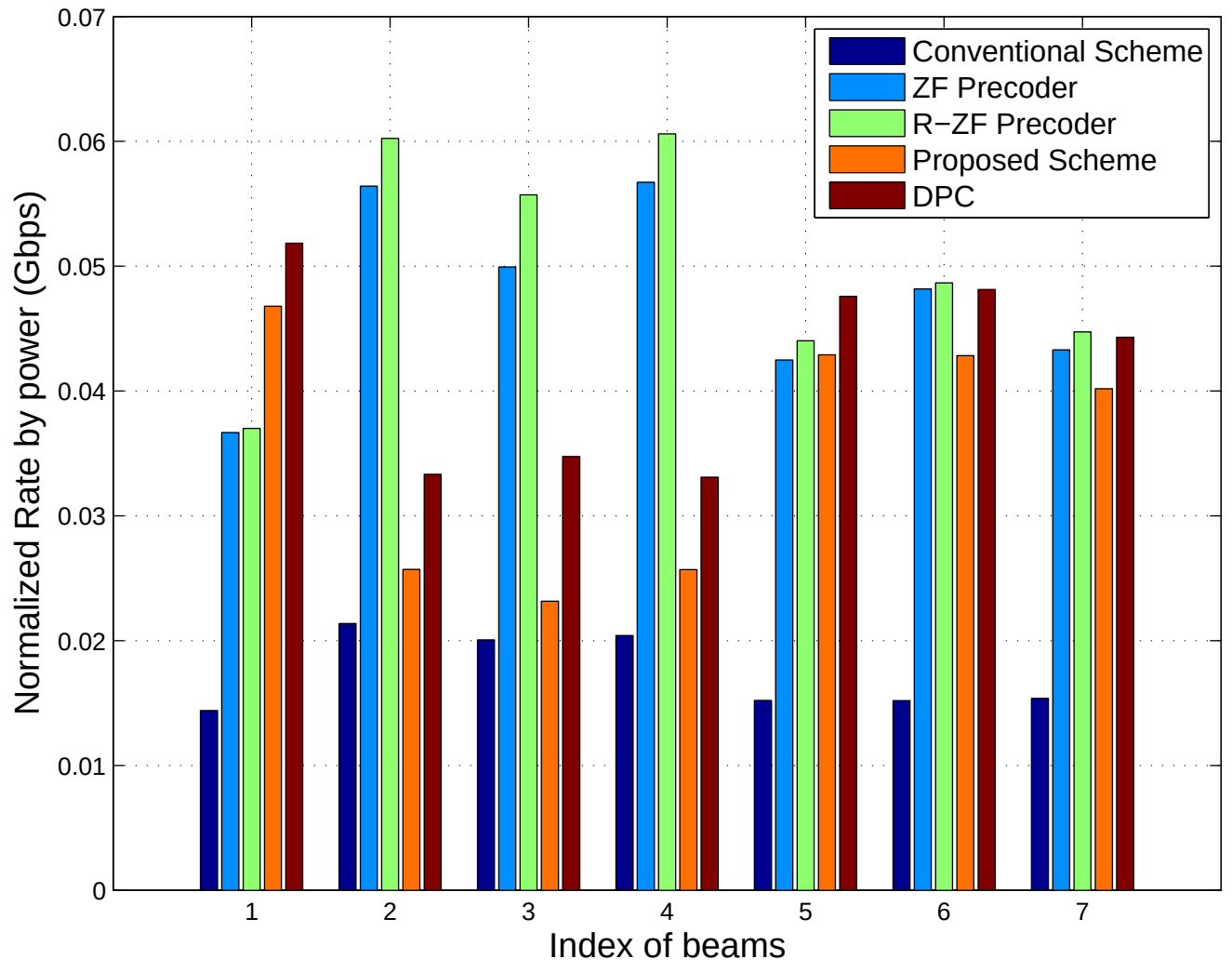

Fig. 4. Comparison of rate normalized by beam power for different schemes.

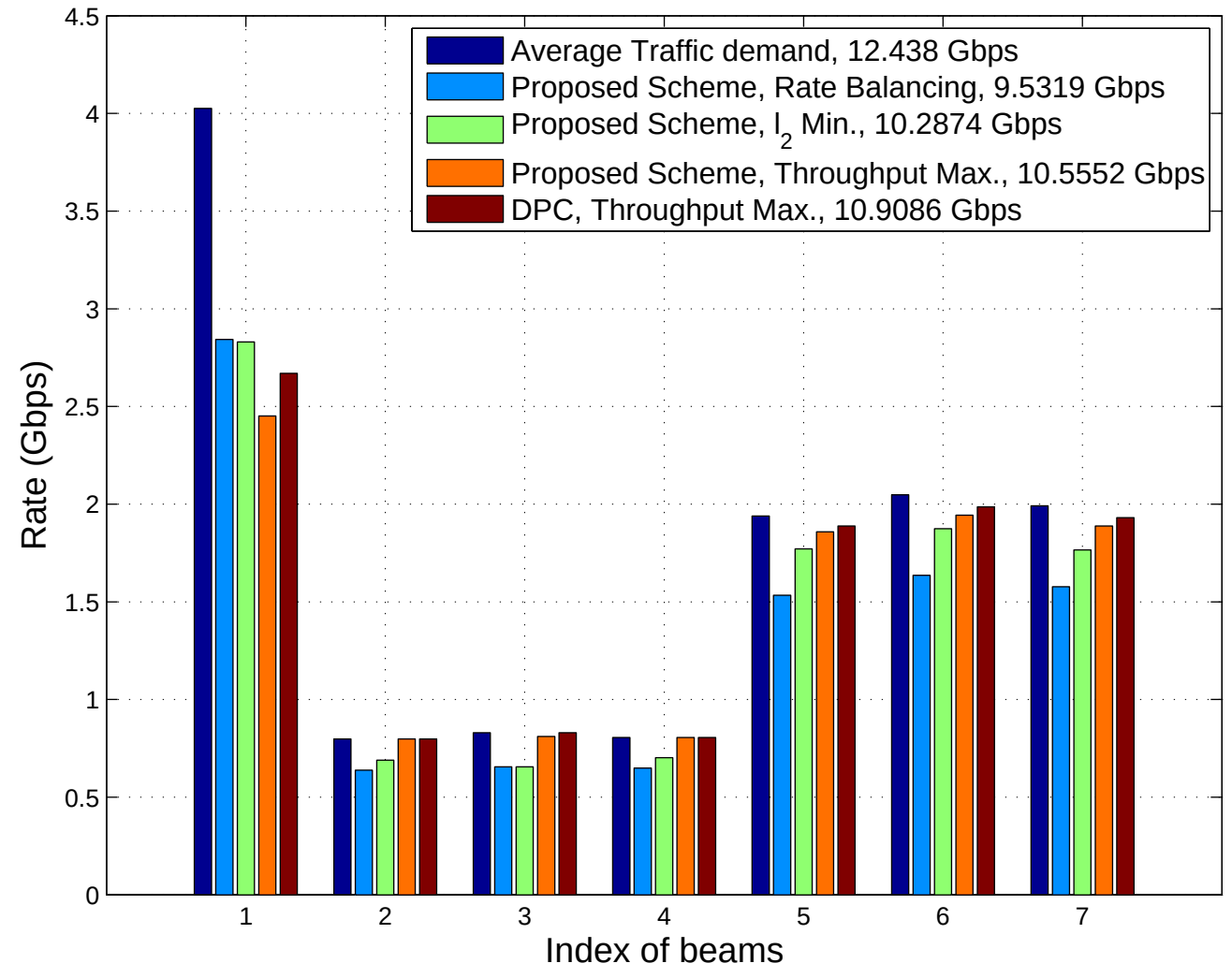

Fig. 5. Comparison of rate with different objectives. In the legend, the name of each scheme is followed by the throughput.

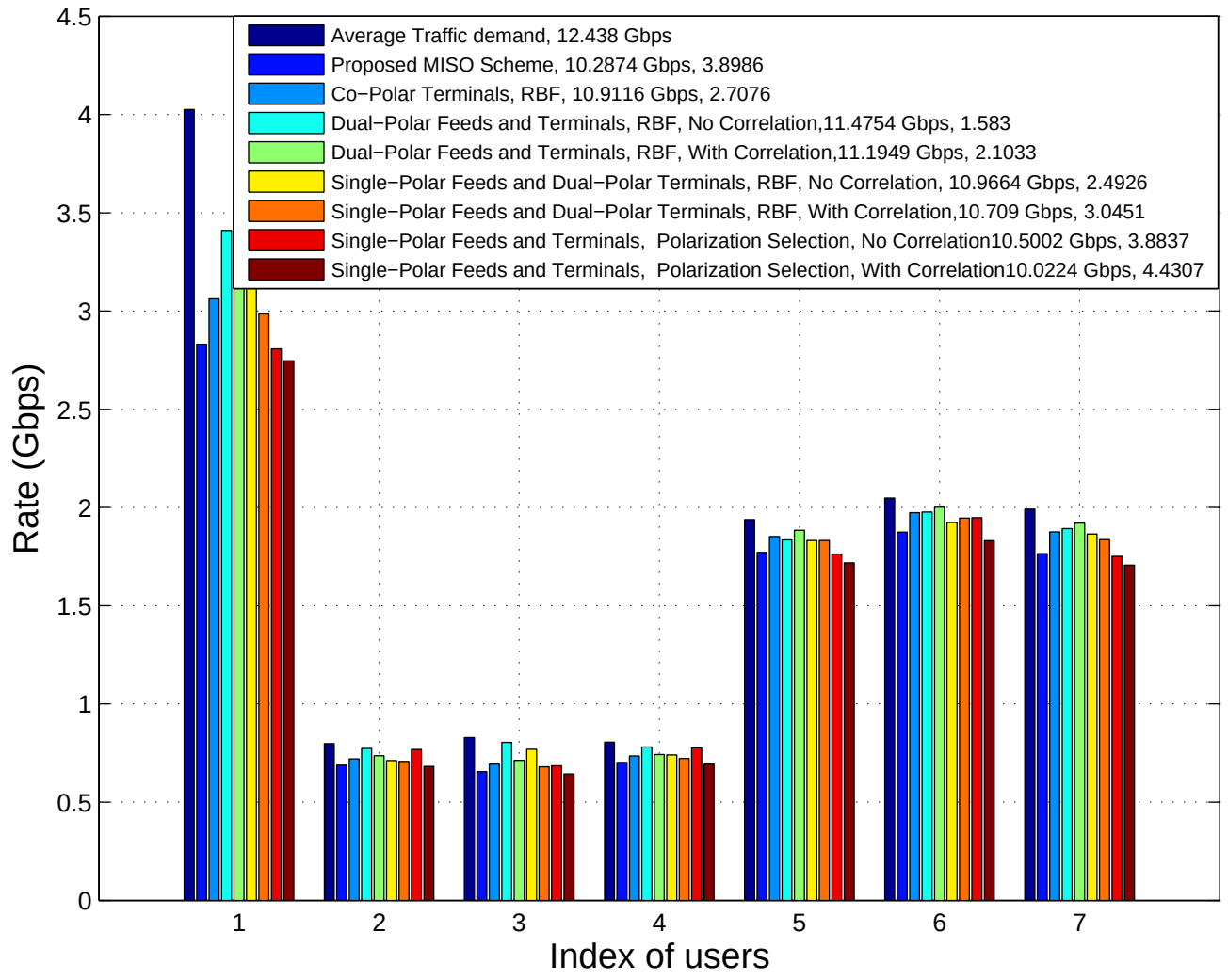

Fig. 6. The effects of co-polarization and dual-polarization antennas with correlation coefficient of 0.9 for different schemes. In the legend, the name of each scheme is followed by the throughput and the average value of the $l_2$ norm.